%% file: ms.tex
\documentclass[sigchi-a]{acmart}

\copyrightyear{2019}
\acmYear{2019}
\setcopyright{none}
\acmConference[CHI'19 Extended Abstracts] {CHI Conference on Human Factors in Computing Systems Extended Abstracts}{May 4--9, 2019}{Glasgow, Scotland UK}
\acmPrice{}
\acmISBN{}
\acmDOI{10.1145/3290607.3312823}
\title{Minimalistic Explanations: Capturing the Essence of Decisions}

\author{Martin Schuessler}
\affiliation{
  \institution{Technische Universit\"at Berlin\\
               Weizenbaum Institute}
  \city{Berlin} \postcode{10623} \country{Germany}}
\email{schuessler@tu-berlin.de}

\author{Philipp Wei\ss{}}
\affiliation{
  \institution{Technische Universit\"at Berlin\\
               Weizenbaum Institute}
  \city{Berlin} \postcode{10623} \country{Germany}}
\email{philippweiss@mailbox.tu-berlin.de}

\begin{document}

\begin{abstract}
  \input{abstract.tex}
\end{abstract}

\maketitle

\input{intro.tex}
\input{method.tex}
\input{results.tex}
\input{discussion.tex}
\input{future_work.tex}

\newpage

\begin{sidebar}
  \vspace{5mm}
  \textbf{Acknowledgements: }
Funded by the German Federal Ministry of Education and Research (BMBF) - NR 16DII113.
During her fellowship at the Weizenbaum Institute, Stefania Druga provided helpful comments on this work.
Berit Wiegmann helped to refine the infographics.
We are also grateful to the anonymous reviewers for their valuable suggestions to mature the ideas presented in this paper.
\end{sidebar}

\input{ms.bbl}


\end{document}

%% file: abstract.tex
The use of complex machine learning models can make systems opaque to
users. Machine learning research proposes the use of post-hoc
explanations. However, it is unclear if they give users insights into
otherwise uninterpretable models. One minimalistic way of explaining
image classifications by a deep neural network is to show only the areas
that were decisive for the assignment of a label. In a pilot study, 20
participants looked at 14 of such explanations generated either by a
human or the LIME algorithm. For explanations of correct decisions, they
identified the explained object with significantly higher accuracy
(\(75.64~\%\) vs. \(18.52~\%\)). We argue that this shows that
explanations can be very minimalistic while retaining the essence of a
decision, but the decision-making contexts that can be conveyed in this
manner is limited. Finally, we found that explanations are unique to the
explainer and human-generated explanations were assigned \(79~\%\)
higher trust ratings. As a starting point for further studies, this work
shares our first insights into quality criteria of post-hoc
explanations. 

%% file: intro.tex
\section{Introduction}\label{introduction}

The impact of machine learning on our society is growing as it is
becoming an integral part of many computer programs. Unfortunately,
systems like deep neural networks that have significantly promoted the
revival of machine learning research are inherently uninterpretable due
to their sub-symbolic nature. Hence researchers are faced with a
fundamental technical barrier to transparency as they have limited
understanding of what these systems are learning and are unable to prove
that they will work on unseen problems \citep{lipton2018}. Nevertheless,
transparency and explainability are an integral component of ethically
aligned design
\citep{theieeeglobalinitiativeonethicsofautonomousandintelligentsystems2017, acmu.s.publicpolicycouncil2017}.
Consequently, interpretable machine learning research has seen a surge
in interest and publications with two main streams of research: The
first suggest new ``simpler'' models that are mathematically more
interpretable yet exhibit comparable performance to uninterpretable
models. The second seeks to explain black-box model predictions with
post-hoc explanations without uncovering the mechanism behind them
\citep{lipton2018}. The running hypothesis that motivates such research
is that displaying explanations can help novice and expert users to
develop trust into a model \citep{ribeiro2016}.

However, there is minimal consensus on a definition for interpretability
\citep{doshi-velez2017, lipton2018} and scholars have argued that
research in this field needs to build more strongly on research on
explanation in philosophy, psychology and cognitive science
\citep{miller2017a}. Furthermore, human factors and real-world usability
aspects are often neglected when new approaches are proposed, which may
be because current interpretable machine learning research is relatively
isolated from HCI research \citep{abdul2018}.

However, interaction with intelligent systems and agents is a
traditional field of HCI. For example, Kulesza et al.
\citep{kulesza2015} introduced \emph{Explanatory Debugging Systems} that
explain their decisions and incorporate user feedback, which was shown
to lead to better predictions, sounder mental models and higher user
satisfaction. Since their implementation has been limited to simple
Naïve Bayes classifiers, these principles and findings may not translate
to complex deep learning models. More recent work from our community
includes work by Binns et al. \citep{binns2018} studying how different
presentation styles of explanation influence justice perception or work
by Rader et al. \citep{rader2018} studying how explanations of the
Facebook news feed algorithm influence the beliefs and judgments.

In this work, we add to this body of research by investigating if
minimalistic post-hoc explanations can capture the essence of a decision
and if they align with human intuition.

%% file: method.tex
\input{figures/info_graphic_data.tex}

\section{Method}\label{method}

A ``full'' explanation of a complex model is often not feasible or even
understandable for humans, which is why explanations need to be
selective in the causes they present \citep{miller2017a}. For the
machine learning task of image classification where an image is assigned
one of several possible labels, \textbf{anchors} are one possible way of
providing such minimalistic explanations. An anchor is the reduction of
the input image to the regions that supported the assignment of a label.
In our pilot study, we compared algorithmically generated anchors to the
gold standard of human explanations. For this purpose, we photographed
several everyday objects and generated anchors for them algorithmically
and manually.

\subsection{Algorithmically Generated
Anchors}\label{algorithmically-generated-anchors}

To generate anchors algorithmically we used the Keras framework
\citep{chollet2015keras} with tensorflow
\citep{tensorflow2015-whitepaper}. We predicted a label for each photo
using the \emph{Inception v3 model}
\citep{DBLP:journals/corr/SzegedyVISW15} trained with the 1000 class
ImageNet training data (Figure \ref{fig:info:data} - Step 1). For the
post-hoc explanation method, we restricted our experiment to
\emph{local interpretable model-agnostic explanations}, generated with
the \textbf{LIME algorithm}. This algorithm was developed by Ribeiro
et al.  \citep{ribeiro2016} in 2016. In a user study, they also
demonstrated its ability to support users in identifying
generalisation error and skewed datasets.

\input{figures/info_graphic_experiment_1.tex}
\input{figures/info_graphic_experiment_2.tex}

For a decision, LIME creates a sparse, linear model \(g\) with
super-pixels as input. The resulting model is interpretable for two
reasons: Firstly, the domain of \(g\) is a super-pixel representation of
the image, which is meaningful for a human. Secondly, the sparsity
constraint enforces that just a few of all super-pixels contribute to
the classification by \(g\), creating a very selective model. The anchor
is obtained by reducing the input image to pixels that supported the
decision (Figure \ref{fig:info:data} - Section B2). Anchors generated in
this fashion can exhibit some rough edges which we smoothed manually. It
is important to note here that different model architectures (e.g.,
vgg16) produce different anchors and how the architecture influences the
anchors is an open research question.

\subsection{Manually Generated
Anchors}\label{manually-generated-anchors}

We showed photos of seven everyday objects to four volunteers recruited
within our institute and asked them to assign a label to the image
(Figure \ref{fig:info:data} - Step 1). Next, we instructed them to mark
up regions of the image that they considered most relevant for their
decision (Figure \ref{fig:info:data} - Step 2). If in doubt explainers
were instructed to consider what regions they considered essential in
such a way that their removal would make it much harder to identify the
object. Finally, their selections were cut out from paper and glued back
to paper smoothing the edges if necessary. Once we had created a couple
of anchors in this fashion, they appeared to be considerably different
from the algorithmically generated ones.

\subsection{Study Design}\label{study-design}

If anchors are selective in a human-understandable way, they should
reduce an image to the essential parts. If this is the case, humans
should be able to identify the object for which an anchor was
generated if the anchor was generated for the correct object label. We
hosted a pilot study with twenty participants, researchers from
multiple disciplines, at the Weizenbaum Institute. In the first half
participants were individually presented with seven anchors of the
seven objects, randomly either algorithmically or manually created.
In a questionnaire, they were asked to identify the object outlined by
the anchor, give a difficulty rating for this task (five-point Likert
scale) and select whether they think the anchor was generated by a
human or by an algorithm (Figure \ref{fig:info:exp1}). In the second
part, we showed participants the original images of the object along
with the anchors they had already seen and the ones they had not
seen. Hence a manually and an algorithmically generated anchor were on
display for each object.  We also marked the anchors that explained a
wrong label. In the questionnaire, we asked participants once again to
determine for each anchor if a human or an algorithm generated
it. Lastly, assuming the anchor had been generated by an algorithm
they were asked to rate the likelihood that they trusted the
underlying classifier to classify objects of the same type correctly
in the future (Figure \ref{fig:info:exp2}).

%% file: figures/info_graphic_data.tex
\begin{marginfigure}
  \includegraphics[width=\marginparwidth]{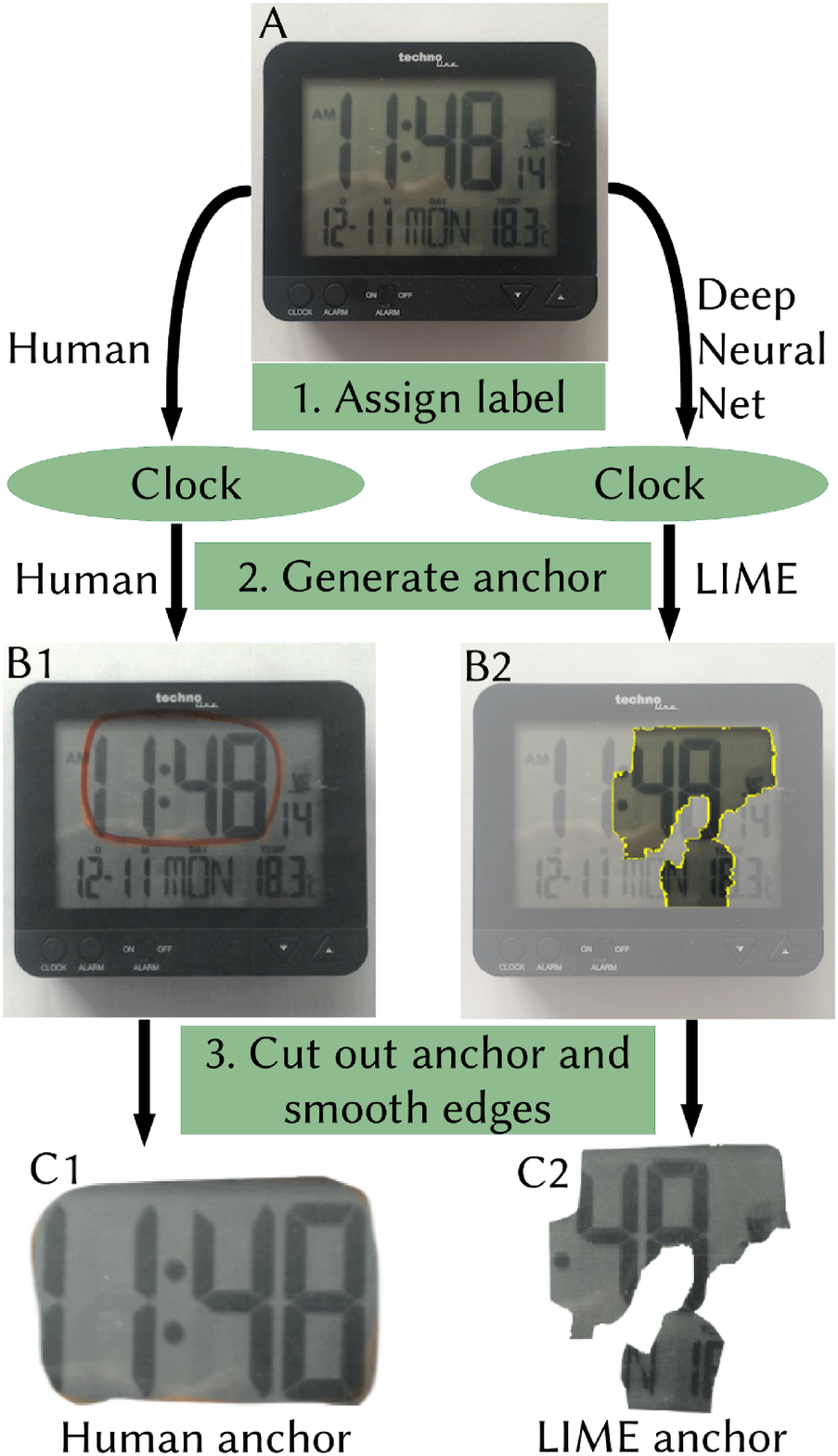}
  \Description{An infographic illustrating how anchors where generated. The illustration uses the photo of a digital watch as an example. This photo is shown at the top. From there the illustration splits between a left-hand side and a right-hand side. On the left-hand side, the anchors are generated with humans. On the right-hand side, anchors are generated with algorithms. Both procedures go through three steps. The first step is the assignment of a label to the image. The second step is the generation of the anchor. The third step is the cutting of the anchor and the smoothening of the anchor’s edges. On the left side, in the first step, the photo is shown to a human who assigns the label “clock” to it. On the right-hand side, the image is fed into a deep neural network assigning the same label. In the next step, the anchor generation step, a human has used a red pen to circle a region on the picture of the clock. In this case, the human had decided to circle the large digital numbers shown by the clock display. On the right-hand side, the anchor was generated using the LIME algorithm. The algorithm calculated which regions where most influential for the assignment of the label "clock". The region is quite different from the one the human selected. The shape is hard to describe but I will try my best. Imagine the shape of Southern Europe from Paris to Croatia. This shape is mapped to the digital clock in such that it covers the digital letters showing the seconds. From there it spans further downwards and Sicily is located above some smaller digital numbers which indicate the date. However, besides these minor differences, the overall most notable difference of these two images is that the human selection is a smooth oval shape whereas the LIME selection has many rough and pointy edges like the shores of the Mediterranean sea. In the final step, these edges are normalized which means that the human selection becomes a bit more pointy and the region LIME selected becomes smoother. }
  \caption{Anchor generation scheme:
    On the left branch a human assigns a label and highlights the anchor.
    On the right branch, a deep neural network assigns the label, and the \bf{LIME} \cite{ribeiro2016} algorithm creates the anchor.
    Both anchors are printed on paper and cut out by hand to smooth the edges.}
  \label{fig:info:data}
\end{marginfigure}

%% file: figures/info_graphic_experiment_1.tex
\begin{marginfigure}
  \includegraphics[width=\marginparwidth]{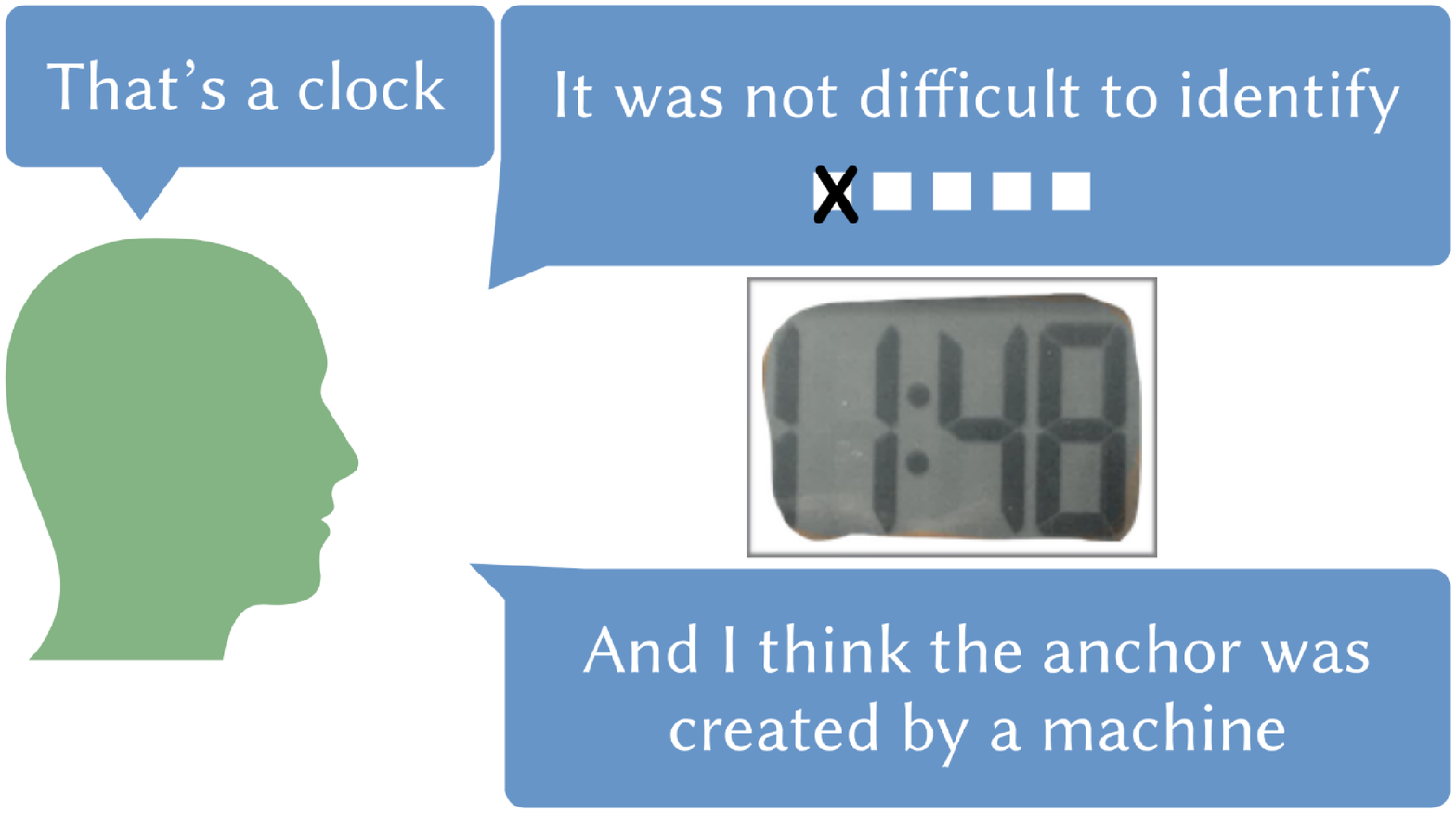}
  \Description{An infographic showing a person looking at 4 digits on a grey oval background. The digits read 11:48. Note that this is the anchor another human selected as illustrated in Figure 1. Speech bubbles above the person illustrate how he or she answers the question in the first stage of the experiment. The first bubble reads “That’s a clock”. The second bubble reads “It was not difficult to identify”. Below this utterance is a 5-point Likert Scale where the first box has been crossed. The third bubble reads “I think the anchor was generated by a machine”}
  \caption{First stage of the experiment:
    For each image one of the two anchors are shown to subjects.
    They decide what the original label was, how difficult it is to recognise the label and finally how the anchor was created.}
  \label{fig:info:exp1}
\end{marginfigure}

%% file: figures/info_graphic_experiment_2.tex
\begin{marginfigure}
  \includegraphics[width=\marginparwidth]{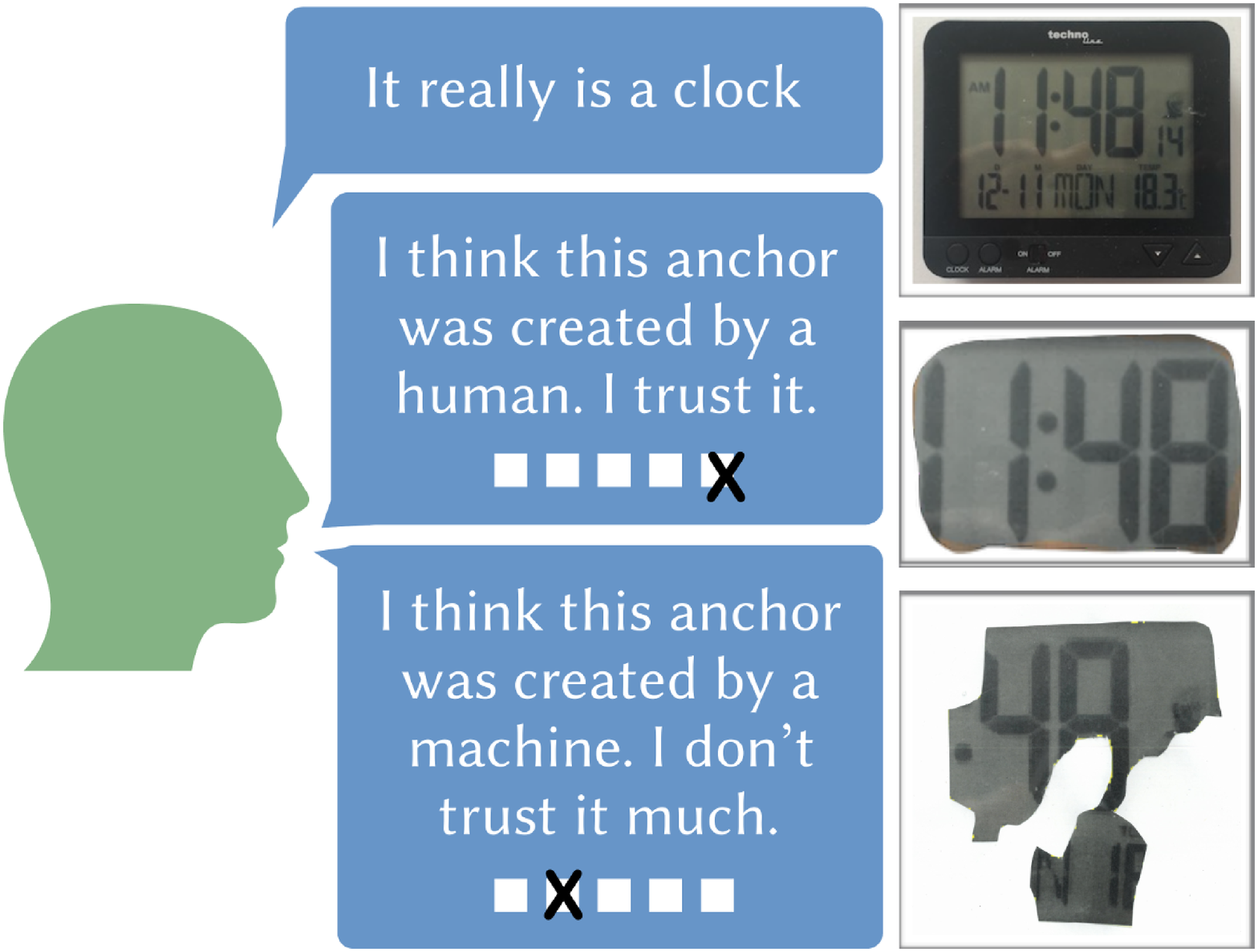}
  \Description{An infographic showing a person looking at three images. The first image is the original photo of the digital watch. About this image, the person is thinking “It really is a clock”. The second image is the anchor generated by a human which was described in Figure 1. About this image the person is thinking: “I think this anchor was created by a human. I trust it.” A Likert scale below has been marked with the highest rating. The third image is the anchor generated by the LIME algorithm which was also described in Figure 1. About this image the person is thinking: “I think this anchor was created by a machine. I don’t trust it much.” A Likert scale below has been marked with a trust rating of two.}
  \caption{The second stage of the experiment:
    The subjects see both anchors and the original image.
    Again they decide which anchor was created by the algorithm.
    They also judge if they would trust the classifier given each explanation and assuming the machine created it.
    }
  \label{fig:info:exp2}
\end{marginfigure}

%% file: results.tex
\section{Results}\label{results}

\input{figures/plot.tex}

Fifteen out of twenty participants submitted their questionnaire which
was optional. We analysed the data using two-way repeated measurement
ANOVAs and report only significant results in this short work. As shown
in Figure \ref{fig:stat} the recognition rate was significantly lower
for explanations that explained the wrong label (\(18.52\%\) vs.
\(75.64\%\); \(F_{(1,105)}=40.14, p<0.001\)). Similarly, the difficulty
rate was significantly higher (\(M=4.70,SD=0.53\) vs.
\(M=2.66,SD=1.59\); \(F_{(1,99)}=43.0754, p<0.001\)). In the first part
of the experiment participants were able to distinguish between
algorithmically and manually generated anchors with an average accuracy
of \(57.45~\%\) which increased to \(82.52~\%\) in the second part where
anchors where displayed pairwise along with the original image. If an
anchor explained an incorrect label, trust ratings were significantly
lower as when it explained the correct label (\(M=2.17, SD=1.05\) vs.
\(M=3.89, SD=1.09\); \(F_{(1,205)}=82.45, p<0.001\)) and participants
trusted manually generated explanations significantly more than
algorithmically generated ones (\(M=3.83, SD=1.25\) vs.
\(M=2.99, SD = 1.29\); \(F_{(1,205)}=6.90, p=0.009\)).

%% file: figures/plot.tex
\begin{marginfigure}
  \includegraphics[width=\marginparwidth]{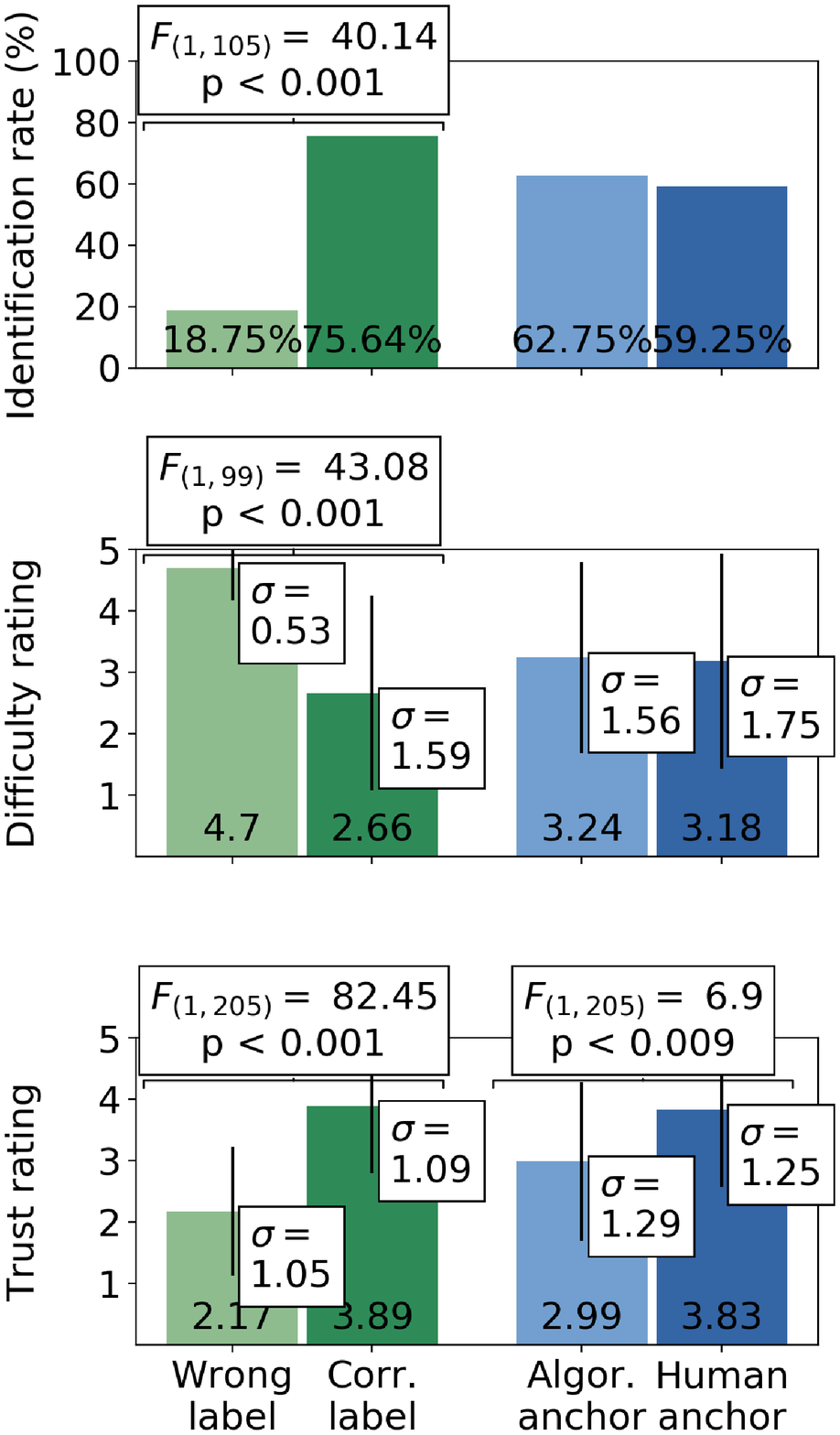}
  \Description{Study results in grouped bar charts. In total there are three charts. Each is showing four bars. The first two bars always compare the two conditions “wrong label” and “correct label”. The last two bars always compare the condition “algorithmically generated anchor” and “human anchor”.
The first bar chart shows the Identification rate in percent. For the wrong label condition, it is 18.75 percent. For the “correct label” condition, it is 75.64 percent. A significant effect exists between the two conditions with F equals 40.14 p smaller than 0.001. For the “algorithmically generated anchor” condition the value is 62.75 percent and for the “human anchor” condition it is 59.25 percent. There is no effect.
The second bar chart shows difficulty rating on a scale from zero to five. For the wrong label condition, it is 4.7 and the standard derivation is 0.53. For the “correct label” condition it is 2.66 and the standard derivation is 1.59. A significant effect exists between the two conditions with F equals 43.08 and p smaller than 0.001. For the “algorithmically generated anchor” condition the rating is 3.24 with a standard derivation of 1.56. The rating for the “human anchor” condition is 3.18 with a standard derivation of 1.75. There is no effect.
The third and last bar chart shows trust ratings on a scale from zero to five. For the wrong label condition, it is 1.05 and the standard derivation is 1.05. For the “correct label” condition it is 3.89 and the standard derivation is 1.09. A significant effect exists between the two conditions with F equals 82.45 and p smaller than 0.001. For the “algorithmically generated anchor” condition the rating is 2.99 with a standard derivation of 1.29. The rating for the “human anchor” condition is 3.83 with a standard derivation of 1.25. A significant effect exists between the two conditions with F equals 6.9 and p smaller than 0.009.}
\caption{Study results.
  The three graphs compare different metrics for anchors of correctly and incorrectly labeled images (left), as well as anchors generated by humans and algorithms (right).
  Top: Identification rate of the correct object label.
  Middle: Difficulty rating of identifying the object.
  Bottom: Trust in the classifier’s decision.
}
  \label{fig:stat}
\end{marginfigure}

%% file: discussion.tex
\section{Discussion}\label{discussion}

In our pilot study participants were able to identify the original
object more accurately and with more ease when an anchor explained the
right label. Hence, in most cases, \textbf{anchors seemed to reduce
images to their essential parts for a given label while being very
selective}. Nevertheless, an identification rate of \(75.64~\%\) is
still leaving room for improvement. In future studies, we plan to allow
participants to reveal additional regions interactively, which could
identify important regions that had been left out by the explainer. Such
feedback data could be used to improve or debug the classifier.

We also found that \textbf{explanations were unique to the explainer}
(human subject or machine learning model respectively) and therefore
considerably different from one another (i.e., anchors C1 and C2 in
Figure \ref{fig:info:data}). Hence it was easy for participants to
distinguish between them once they were displayed side by side. Some
participants mentioned that they saw a pattern in how they differed,
stating that humans are more focused on the objects overall shape and
the co-occurrence of region whereas the algorithm focussed on
object-specific patterns in sub-regions. They also trusted the manually
created anchors significantly more (3.89 vs.~2.17). Whether this is due
to a general tendency to trust humans more is left to be investigated.
Interestingly participants mentioned that they did not expect
explanations to overlap or to be similar, but they expected them to
align with their intuition. This shows that there can be more than one
reasonable explanation for a given decision.

When creating anchors manually, participants often circled different
regions that were overlapping or connected stating that the occurrence
of both regions together or in a particular spatial arrangement is what
made them assign a specific label (see Figure \ref{fig:key}). However,
mapping such an explanation to a set of sub-regions is not possible.
Hence, \textbf{anchors can only communicate very few reasons for a given
decision}. Future research could consult expertise from cognitive
psychology and social science \citep{miller2017a} about how humans
generate and look at explanations. Such insights can be used to extend
LIME or other post-hoc methods to convey more decision making context
such as the relationships between regions. It is important to mention
here that many interpretable models such as rule-based systems or
classification trees provide explanations for the combination of
features to a decision. Furthermore, explanations are not limited to the
use of input features. Their expressiveness can be enhanced with the use
of other media and modalities (see \citep{lipton2018} for examples).
Sevastjanova et al. \citep{sevastjanova2018} even outlined a very
promising design space for the combination of \emph{verbalisation} and
\emph{visualisation} to produce even richer explanations.

\input{figures/key.tex}

%% file: figures/key.tex
\begin{marginfigure}
  \includegraphics[width=\marginparwidth]{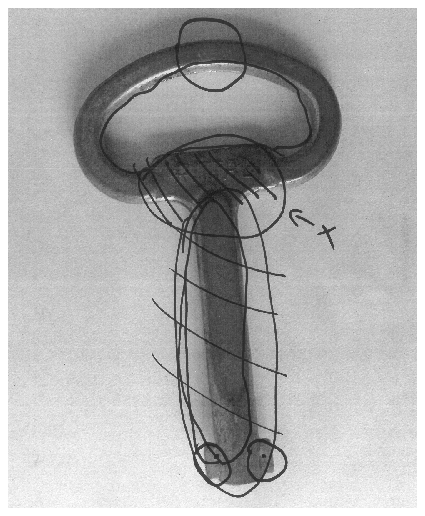}
  \Description{An image showing a bottle opener that is also a key for a door of an electric appliance room. The image has been marked with several regions as described in Figure 1. Almost the entire object is covered with small regions. Two regions have been crossed out to indicate that they are not as important as the other ones.}
  \caption{Participants highlights used to explain why s/he saw a key in this image of a bottle opener. Several circles cover almost the entire object because their arrangement as a whole was considered significant. The hatched area indicates that this region was of lesser importance.}
  \label{fig:key}
\end{marginfigure}

%% file: future_work.tex
\section{Future Work and Conclusion}\label{future-work-and-conclusion}

 We aim to repeat this study with a more thorough design (no convenience
sampling, better isolation of factors, improved shape of anchors,
standardised questionnaires). In this experiment, we studied a very
abstract notion of trust as the faith in a models performance. Following
the argumentation of Doshi-Velez et al. \citep{doshi-velez2017} trust
should instead be evaluated in respect to some real-world desiderata and
more carefully operationalised. For example, one could base the reward
for the experiment on the participant's ability to rely on the system
appropriately. In such an experiment post-hoc explanations could be
compared to real explanations, placebo explanation or simple model
performance statistics. In future studies, we also seek to asses another
quality indicator of explanations: their \emph{decision-contrasting
capabilities} \citep{lipton2018, miller2017a}. Since anchors only
provide information about why a label was assigned, we plan to
investigated if they can also provide useful information about why
another label was not chosen.

In this work, we found that anchors are very minimalistic explanations
that can be very selective. Even though they retain the essence of a
decision, it is worth investigating how they could convey more
decision-making contexts. We see this early work as a starting point for
a series of human grounded evaluations \citep{doshi-velez2017} that
asses the practical interpretability provided by post-hoc explanations
and interpretable models.